\documentclass[11pt]{article}

\usepackage{deauthor}
\usepackage{amsmath, amsfonts}  
\usepackage{epsfig,endnotes,xspace,url,diagbox}

\usepackage{xcolor}

\usepackage{hyperref}
\hypersetup{
    colorlinks=true,       
    linkcolor=tableofcontent, 
    citecolor=citecol,        
    urlcolor=black,           
}

\usepackage{bbm}
\usepackage{enumerate}
\usepackage{varioref}
\usepackage{graphicx}
\usepackage{subfigure}

\usepackage{caption}
\usepackage{float}
\usepackage{cleveref}

\usepackage{epstopdf}

\usepackage{footnote}

\usepackage{colortbl} 

\usepackage{multirow}
\usepackage{comment}
\usepackage{color}
\usepackage[utf8]{inputenc}
\usepackage{mathtools}

\usepackage{wrapfig}
\usepackage{tabularx}
\usepackage{xspace}

\usepackage{algorithm}
\usepackage{algorithmic}

\usepackage{cite}
\usepackage{booktabs}
\usepackage{tcolorbox}
\usepackage{enumitem}
\usepackage{url}

\definecolor{citecol}{RGB}{0,0,255} %
\definecolor{tableofcontent}{RGB}{0,0,0} 

\setlist[itemize]{leftmargin=*,noitemsep, topsep=0pt}

\newcommand{\ie}{\emph{i.e., }}
\newcommand{\eg}{\emph{e.g., }}

\newcommand{\mypara}[1]{\smallskip\noindent\textbf{#1.} \xspace}

\usepackage{todonotes}

\begin{document}

\title{Beyond Data Privacy: New Privacy Risks for Large Language Models}

\author{Yuntao Du$^\dagger$ ,\quad Zitao Li$^{\ddagger}$,\quad Ninghui Li$^\dagger$,\quad Bolin Ding$^{\ddagger}$ \\
    $\dagger$ Department of Computer Science, Purdue University\\
     \texttt{\small \{ytdu, ninghui\}@purdue.edu}\\
    $\ddagger$ Alibaba\\
     \texttt{\small \{zitao.l, bolin.ding\}@alibaba-inc.com}\\
    }

\maketitle

\begin{abstract}
Large Language Models (LLMs) have achieved remarkable progress in natural language understanding, reasoning, and autonomous decision-making. 
However, these advancements have also come with significant privacy concerns. 
While significant research has focused on mitigating the data privacy risks of LLMs during various stages of model training, less attention has been paid to new threats emerging from their deployment. 
The integration of LLMs into widely used applications and the weaponization of their autonomous abilities have created new privacy vulnerabilities.
These vulnerabilities provide opportunities for both inadvertent data leakage and malicious exfiltration from LLM-powered systems. 
Additionally, adversaries can exploit these systems to launch sophisticated, large-scale privacy attacks, threatening not only individual privacy but also financial security and societal trust.
In this paper, we systematically examine these emerging privacy risks of LLMs. 
We also discuss potential mitigation strategies and call for the research community to broaden its focus beyond data privacy risks, developing new defenses to address the evolving threats posed by increasingly powerful LLMs and LLM-powered systems.

\end{abstract}

\section{Introduction}

Recent advancements in deep learning, particularly in natural language processing, have led to the development of large language models (LLMs).
Over the past few years, LLMs have demonstrated impressive capabilities in understanding and generating human language. 
These models are rapidly growing in size and effectiveness, yielding breakthroughs and attracting increasing research and social attention. 
Beyond natural language understanding, their emergent abilities~\cite{arxiv22emergent} have enabled them to achieve unparalleled performance on complex tasks. 
As a result, LLMs are no longer standalone models but are increasingly integrated as core decision-making components in larger systems, such as interactive chatbots~\cite{gpt4,claude,arxiv24qwen} and autonomous agents~\cite{agents_anthropic,arxiv24sweagent,uist23agent}.

However, this rapid development comes with growing concerns about its privacy implications.
As a primary source of privacy risk, LLMs are trained on vast, internet-scale corpora that often contain sensitive personal information and copyrighted content. 
These data privacy risks are amplified when models are fine-tuned on private, proprietary datasets.
Studies~\cite{acl23finetunellmmia,nips24datasetinference,ccs24janus} have shown that LLMs can memorize and inadvertently leak training data across various model learning stages, raising issues related to training data extraction~\cite{usenix21extracting}, copyright infringement~\cite{nips24llm-copyright}, and test set contamination~\cite{iclr24contamination}.

Beyond the risks of training data leakage, privacy threats also emerge from the integration of LLMs into larger, more complex systems, which we refer to as LLM-powered systems. 
These systems, especially those that use LLMs as decision-making engines in agent-based applications~\cite{cursor, wang2024agentsurvey}, introduce new vulnerabilities and expand the potential attack surface for privacy violations. 
For instance, a user may share personal information with an LLM-based chatbot in order to receive personalized responses or suggestions. 
However, this information could be exfiltrated through side channels~\cite{carlini24timing} or unintentionally disclosed by the model itself~\cite{iclr24beyond}. 
Such risks are not inherent to the LLM alone but emerge from the architecture of interactions between users, models, and other system components.
As LLM-powered applications become increasingly widespread in both daily life and professional domains, these privacy risks become more prominent and urgent.

A third category of privacy threats arises from the advanced reasoning and autonomous decision-making capabilities of LLMs, which create new opportunities for malicious exploitation.
These capabilities enable adversaries to automate sophisticated attacks at unprecedented scale and speed, substantially lowering the barrier to entry for cyberattacks.
For instance, an attacker could instruct an LLM to infer sensitive attributes, such as a user’s demographics, from their public online posts, leading to de-anonymization and other severe cybercrimes~\cite{arxiv25autoprofiler,usenix24malla}. 
Similarly, LLMs can be leveraged to launch large-scale, highly personalized social engineering campaigns, resulting in significant financial and societal consequences~\cite{se_example}. 

\begin{figure}[t]
    \centering
    \includegraphics[width=0.99\textwidth]{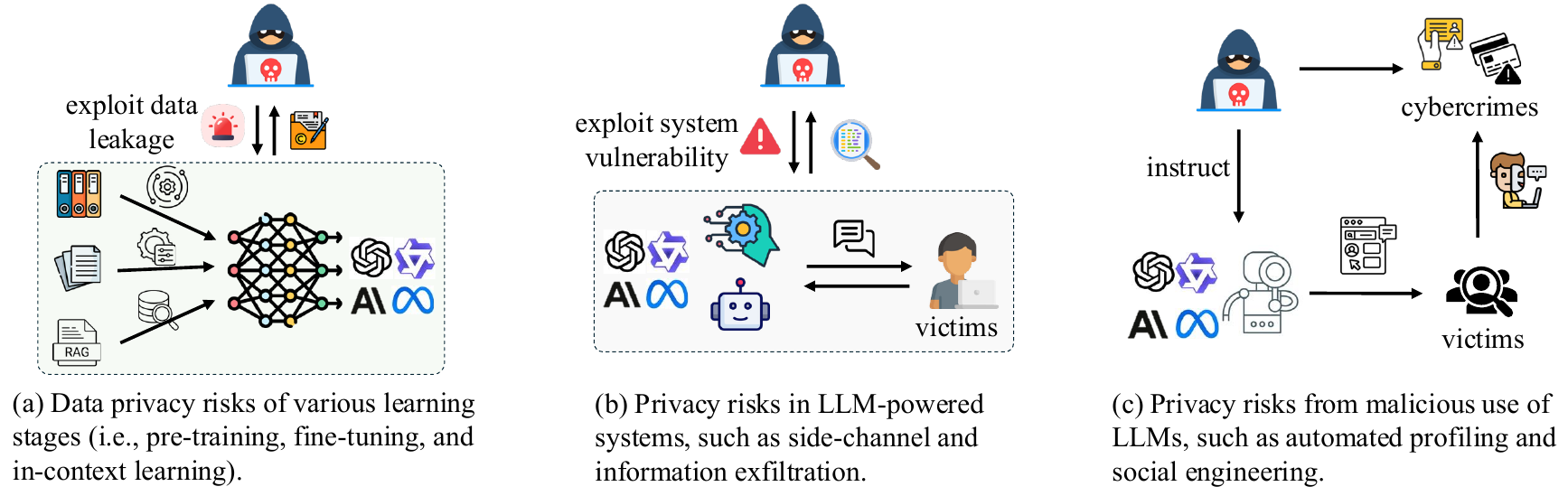}
    \caption{Illustrations of different types of privacy risks posed by large language models.}
    \label{fig:illustration}
\end{figure}

\mypara{Positioning and Contribution}
Existing research has predominantly focused on training data privacy issues of LLMs. 
In contrast, less attention has been paid to the privacy threats posed by LLM-powered systems and the malicious use of LLMs.
The privacy risks arising from LLM-powered systems and their malicious use represent a paradigm shift. 
These risks are not rooted in sensitive training data, but in the increasingly powerful autonomy of LLMs. 
As a result, existing data privacy frameworks may not always be well-suited to analyze or mitigate these emerging threats.
This paper aims to bridge this gap by providing a comprehensive study of the new threat landscape introduced by LLMs. 
We systematically examine the privacy risks that arise from these models and their applications, and discuss potential mitigation strategies, calling for research efforts and greater public awareness to address these emerging privacy challenges.

\mypara{Related Work}
Several studies~\cite{weidinger2021ethical,li2025rethinking, dritsas2024survey, arxiv23privacysurvey, vldb24llm-pbe, arxiv23llmprivacy} have surveyed the data privacy risks of LLMs and explored mitigation strategies, 
However, these studies do not cover the emerging privacy threats posed by the increasing integration of LLM-powered systems and the potential malicious use of LLMs, which we identify as critical new privacy risks.
Another line of research~\cite{chi24privacysurvey, fate22privacyllmsurvey,xie2025evaluating,chen2025clear} has examined the privacy implications of LLMs during user interactions.
Our work builds on and extends these prior studies, providing a systematic and comprehensive analysis of the privacy risks posed by LLMs.


\mypara{Roadmap}
In~\Cref{sec:background}, we introduce the necessary background on LLMs and their key developmental trends. 
In~\Cref{sec:data}, we discuss the primary data privacy risks associated with LLMs at different stages of training.
In~\Cref{sec:system}, we analyze the privacy threats of LLM-powered systems and discuss potential mitigations. 
In~\Cref{sec:use}, we examine the privacy threats arising from the malicious use of LLMs. 
Finally, we conclude and discuss future directions in~\Cref{sec:conclusion}.

\section{Background on Large Language Models}
\label{sec:background}

A language model (LM) is a type of machine learning model for natural language processing. 
In general, an LM estimates the generative likelihood of sequences of words by predicting the probabilities of future or missing tokens\footnote{A token refers to the smallest semantic unit processed by the model, which can be a character, subword, or word.}. 
In recent years, large language models (LLMs), trained on massive datasets for token prediction, have achieved unprecedented performance across a wide range of applications~\cite{arxiv23llm-survey}. 
The scaling of these models has also unlocked remarkable emergent abilities not present in their smaller counterparts~\cite{arxiv22emergent}, including in-context learning~\cite{arxiv22icl_survey}, analogical reasoning~\cite{nature23emergent}, and the capacity to power autonomous agents~\cite{nips23mind2web}.

With the rapid development of LLMs, LLM agents have emerged and become increasingly popular. 
These are intelligent entities powered by LLMs that are capable of autonomously carrying out complex tasks—such as conducting in-depth research or managing computer operations—while adapting to specific user needs~\cite{wang2024agentsurvey}.
This shift represents not only a major technological advancement but also a reimagining of human–machine interaction. 
Their impressive capabilities have already been applied in a wide range of domains, from chatbots~\cite{chatgpt} to professional tools like programming assistants~\cite{cursor}.

\mypara{Trends on LLMs} 
Despite these advancements, the rapid evolution of LLMs also introduces new privacy risks. 
We identify three key trends that are closely related to growing privacy challenges:

\begin{itemize}
    \item \textit{Trained on Sensitive Data.} 
    LLMs are trained on vast amounts of diverse data, which may include sensitive or copyrighted information. 
    Moreover, the advanced capabilities of these models also enable them to access and utilize sensitive data through fine-tuning or in-context learning, especially when building proprietary models or personalized LLM-based applications~\cite{arxiv24personalllm}. 
    This further increases the potential privacy risk of sensitive data that is trained on LLMs.    
    \item \textit{Incorporating into Popular Applications.} 
    LLMs are increasingly embedded into core components of widely used software and professional tools. 
    For example, they serve critical roles in domains such as code generation for software development, document analysis in legal and medical contexts, and as reasoning engines for autonomous agents that interact with external systems and platforms.
    As LLMs become ubiquitous in these daily applications, the surface area for privacy risks expands significantly, as adversaries can exploit these models in a wide range of sensitive contexts.
    \item \textit{Growing Capability and Accessibility.}  
    LLMs are rapidly evolving beyond text-only capabilities. 
    Recent advances in vision-language models (VLMs) enable multimodal reasoning over both text and images~\cite{gpt4v,arxiv23gemini}.  
    At the same time, access to powerful commercial and open-source models has become dramatically easier and much more affordable~\cite{arxiv25llm-report}.  
    This combination of greater capability and accessibility empowers adversaries, giving them opportunities to exploit these advanced, autonomous systems to perform privacy-infringing attacks.
\end{itemize}

Taken together, these trends create a new privacy landscape that both amplifies data privacy concerns and introduces new privacy threats. 
The following sections will analyze these risks in detail, exploring their implications and potential mitigation strategies.


\section{Data Privacy Risks in LLMs}
\label{sec:data}


LLMs themselves pose a significant privacy risk, as they, like other machine learning models, have been shown to memorize elements of their training data~\cite{usenix21extracting,iclr22memorizationllm}.
These data privacy risks can arise at multiple stages of the model learning process.
From a parametric training perspective, LLMs typically undergo both a large-scale \emph{pre-training stage} and subsequent \emph{fine-tuning stages}. While LLMs are pre-trained and fine-tuned on massive corpora, and most state-of-the-art systems~\cite{gpt3,gpt4,llama3,arxiv24qwen,claude} do not disclose the provenance of their training data, concerns regarding potential privacy breaches have become increasingly pronounced.
Moreover, due to their generative nature, LLMs support \emph{in-context} learning, which provides a simple yet powerful mechanism for adapting inference behavior without modifying model parameters. 
These unique characteristics not only enhance the utility of LLMs but also broaden the avenues through which sensitive information may be exposed.
In the following discussion, we examine two of the most prominent privacy threats to LLMs, membership inference and training data extraction, along with their privacy implications and potential enhancements.

\subsection{Membership Inference Attack}
Closely connected with Differential Privacy (DP)~\cite{dwork_dp,ccs13membership}, Membership Inference Attack (MIA)~\cite{sp17miashokri} has become a widely adopted approach for privacy auditing of machine learning (ML) models~\cite{arxiv20privacy_meter,tensorflow_mia}. 
MIA assesses how much a trained ML model reveals about its training data by determining whether specific query instances (or documents) were included in the dataset.
In the context of LLMs, membership inference attacks have been applied at different stages to explore the privacy risk.

\begin{itemize}
    \item \textit{Pre-training Stage.} 
    Early studies~\cite{usenix24llmmia,iclr24mink,emnlp24llmmia,emnlp24llmdetection} propose various membership signals that are derived from LLM's outputs to distinguish members from non-members.
    However, subsequent studies~\cite{satml25sokllmmia,satml25positionllmmia} have identified fundamental flaws in the evaluation methodologies of prior studies. 
    Particularly, the use of temporal data to separate members from non-members introduces subtle distributional shifts between the two groups, resulting in unreliable attack performance that does not accurately reflect true privacy leakage.
    Under more rigorous setups, existing MIAs~\cite{icml25mink++,arxiv24llmmia,arxiv24llmmiaem,sp25llmmia} perform no better than random guessing when targeting pre-training data of LLMs.
    This ineffectiveness is largely due to the fact that each data point is typically used only once during pre-training~\cite{nips23scaling,llama}, and the vast diversity of training corpora further dilutes the influence of a single example~\cite{duan2024membership,usenix25labelllmmia}. 
    \item \textit{Fine-tuned Stage.}  
    Following the pre-training stage, the fine-tuning stage requires substantially fewer resources and focuses on adapting pre-trained models to domain-specific downstream tasks.
    However, fine-tuning datasets frequently contain personally identifiable information (PII)~\cite{ccs24janus}, copyrighted material~\cite{liu2024rethinking}, or sensitive organizational data~\cite{nist_2024}.
    MIAs against fine-tuned LLMs leverage techniques such as prompt calibration~\cite{nips24spvmia}, hypothesis testing~\cite{emnlp22empirical_adapter}, and ensemble methods~\cite{usenix25soft}.
    Moreover, recent work~\cite{aistats25llmmiaaligment} has shown that human preference data used for alignment tuning (\eg via Direct Preference Optimization (DPO)~\cite{icml24llmdpo}) is also vulnerable to MIAs. 
    Compared to attacks on pre-trained LLMs, MIAs against fine-tuned models are markedly more effective. 
    This increased vulnerability arises because fine-tuning datasets are considerably smaller, and fine-tuning often involves multiple training epochs. 
    These factors increase the model's memorization of sensitive data, posing heightened privacy threats in fine-tuned LLMs.
    \item \textit{In-context Learning Stage.} 
    Fine-tuning LLMs for specific domains involves non-trivial computation via parameter updates.
    In-Context Learning (ICL)~\cite{arxiv22icl_survey} has emerged as a popular, more efficient adaptation paradigm, as it does not require modifying model parameters.
    In ICL, private data are provided as demonstrations within the prompt itself to guide the model's inference for a specific task. 
    These demonstrations can be manually prepared~\cite{gpt3} or dynamically retrieved from a private knowledge base using Retrieval-Augmented Generation (RAG) systems~\cite{lewis2020retrieval, asai2024self, shi2023replug}. 
    The vulnerability of ICL to MIAs has been explored using methods such as prompt injection~\cite{arxiv24ragmia}, analysis of semantic similarity~\cite{arxiv25ragmia}, or measuring contextual influence~\cite{arxiv25ragmia}.
    Since ICL relies on only a few demonstration examples, each one has a significant impact on the model's output and performance, making these attacks highly effective at identifying private data.
\end{itemize}

\mypara{Growing Threats}
While current MIAs have been effective at assessing privacy risks for fine-tuned and in-context data in LLMs, they have shown limited success against pre-trained models. 
However, this does not mean that pre-trained models are free from privacy risks. 
Two emerging research directions make MIAs more powerful, presenting a growing threat to the privacy of pre-training data.
The first direction shifts the focus from analyzing individual data instances to examining larger collections of data. 
Recent studies~\cite{nips24datasetinference} show that instead of detecting membership at the sentence level, aggregating membership signals across multiple sentences within a document can reliably reveal whether that dataset was included in training. 
This has been extended to paragraphs and entire collections, showing that these larger structures are also vulnerable to MIAs~\cite{arxiv24pretrainllmscal}. 
The second direction involves more advanced and computationally intensive attacks. 
Recent research~\cite{hayes2025strong} shows that training hundreds of shadow models~\cite{sp17miashokri} to exploit behavioral discrepancies can significantly enhance MIA effectiveness. 
These stronger attacks, building on prior successes against classifiers~\cite{sp22lira, icml24rmia}, further elevate the risks to pre-trained data privacy.
Together, these emerging research trends indicate that future improvements in MIAs could amplify the privacy risks of pre-trained LLMs.

\mypara{Privacy Implications}
Effective MIAs on LLMs have serious trustworthy risks, such as the leakage of copyrighted content~\cite{icml24llmcopyright,nips24llm-copyright} or test set contamination from evaluation benchmarks~\cite{iclr24contamination}.
Moreover, MIAs can serve as a fundamental building block for more sophisticated attacks, such as training data extraction~\cite{usenix21extracting,usenix23diffusion_extract}, or as a core component in data auditing systems~\cite{vldb24llm-pbe,nips24datasetinference,ccs24dataaudit}. 
The issue is also highly relevant to recent ongoing lawsuits alleging unauthorized data use in model training, such as The New York Times vs. OpenAI~\cite{nyt&openai}, Getty Images vs. Stability AI\cite{stabilityai} and Doe vs. GitHub\cite{doe&github}.

\mypara{Potential Mitigation Strategies}
Machine learning with differential privacy (\eg DP-SGD~\cite{ccs16dpsgd} and PATE~\cite{iclr17pate}) is an effective defense mechanism against privacy attacks, including MIAs.
Studies~\cite{iclr22dpsgdllm, ponomareva2023dp, iclr22dpsgdfinetune} have applied DP-SGD to fine-tune LLMs on sensitive domains, which may degrade model utility under reasonable privacy budgets.
For in-context learning scenarios, various DP-based techniques have been proposed, such as differential private prompt tuning~\cite{duan2023flocks, hongdp} and differential private synthetic text generation~\cite{xie2024differentially, tang2024privacy}.
In addition to the theoretical privacy guarantees that DP offers, many empirical privacy methods~\cite{usenix25soft, lora} have shown effectiveness against MIAs. 
For example, LoRA~\cite{lora}, a widely used efficient fine-tuning method for LLMs, demonstrates better privacy preservation than full fine-tuning.
However, these empirical defenses may be compromised when facing stronger MIAs~\cite{hayes2025strong}, and there remains a notable trade-off between privacy and utility, especially for pre-trained LLMs.

\subsection{Training Data Extraction}

Training data extraction refers to the risk of partially or fully reconstructing samples from the training dataset by interacting with a trained LLM~\cite{usenix21extracting,iclr22memorizationllm,inlg23verbatim,nips24llmmeorization,iclr25extract}.
This threat goes beyond merely identifying whether a particular instance was part of the training set (as in membership inference); it involves recovering the training data itself, posing a more severe privacy threat.
Similar to MIAs, data extraction attacks have been studied at various stages of the LLM training pipeline:

\begin{itemize}
    \item \textit{Pre-training Stage.} Research has shown that large amounts of data can be extracted from GPT-2~\cite{gpt2} by repeatedly querying it with different prompt prefixes~\cite{usenix21extracting}. 
    Additionally, studies~\cite{sp23llmpii, nips23propile} have demonstrated that LLMs can unintentionally leak personally identifiable information (PII), such as a person's full name, address, and phone number. 
    More recently, studies~\cite{naacl25extract} have used a probabilistic extraction approach to successfully recover pieces of copyrighted content, such as excerpts from books, from open-weight models.
    \item \textit{Fine-tuned Stage.}  Unlike the pre-training stage, where most data extraction attacks are conducted in a black-box setting, data extraction during fine-tuning often assumes that both the original and fine-tuned model weights are publicly available. For example, the popular reasoning model DeepSeek-R1~\cite{arxiv25deepseekr1} is trained from DeepSeek-Base~\cite{deepseekv3}, both of which have open weights; however, the data used to train R1 has not been made public. 
    Recent studies~\cite{arxiv25approximte} propose efficient data selection strategies that can identify potential training data from a large data pool by matching the gradients from the base model to the fine-tuned model, demonstrating the feasibility of extracting fine-tuning data.
    \item \textit{In-context Learning Stage.} 
    Studies have shown that it is possible to recover the prompt used for in-context learning via model inversion~\cite{iclr24llminversion} or prompt stealing~\cite{sp25promptsteal, arxiv24promptstealing}. 
    The privacy risks are heightened when using Retrieval-Augmented Generation (RAG) systems for LLMs, where attackers try to extract text data from a private database of RAG models through black-box access. 
    Various attacks have been proposed to extract data from RAG systems, including adversarial prompt injection~\cite{iclr25extractrage,ccs24stealprompt,acl25pig}, agent-based attacks~\cite{arxiv25rageextract}, and backdoor attacks~\cite{arxiv24rageextractbackdoor}.
\end{itemize}

Different approaches have been proposed to evaluate the effectiveness of data extraction attacks. 
The most widely used metric is eidetic memorization (or verbatim extraction)~\cite{usenix21extracting} and its variations~\cite{icml22deduplicate, iclr22memorizationllm, nips22memorization, inlg23verbatim, nips23counterfactual}. 
This metric requires the model to reproduce the memorized data exactly when given an appropriate prompt. 
To relax this strict requirement, other studies propose approximate memorization metrics~\cite{inlg23verbatim, icml22deduplicate}, which measure the string or semantic similarity between the model's output and the training data as memorization efficacy.

\mypara{Privacy Implications}
Data extraction from LLMs raises serious privacy and copyright concerns. 
Extraction not only produces a ``copy'' of training data but also reveals that a model has memorized such data internally. 
This evidence is central to ongoing legal debates about whether training an LLM on copyrighted material constitutes fair use. 
Furthermore, it poses a significant risk to the personal or proprietary data used in LLM-based applications such as RAG systems, underscoring the profound privacy challenges in deploying LLMs.

\mypara{Growing Threats}
Recent studies~\cite{arxiv24cooper, naacl25extract} highlight that since LLMs are probabilistic, the memorization should be assessed probabilistically. 
These studies introduce probabilistic discoverable extraction~\cite{naacl25extract}, which quantifies the likelihood that a model, under a given decoding scheme, will reproduce a verbatim target suffix when prompted with a specific prefix. 
This approach not only refines the understanding of memorization in LLMs but also shows how easily certain pieces of data can be extracted from these models.
Further research~\cite{arxiv25extracting} argues that measuring memorization solely by average extraction rates is insufficient. 
Instead, it should focus on identifying specific pieces of copyrighted or private text that are most likely to be memorized by the model. 
By enhancing extraction methods and pinpointing highly memorized fragments, these approaches increase the efficacy of data extraction, highlighting the potential for more targeted privacy violations.


\section{Privacy Risks in LLM-Powered Systems}
\label{sec:system}

The integration of LLMs as core components in larger, complex systems introduces new vectors for privacy risks.
For example, LLM-based chatbots, such as ChatGPT~\cite{gpt4} and Claude~\cite{claude}, have become the primary interface through which users interact with LLMs. 
In these interactions, users often share personal narratives and sensitive details to seek advice or obtain personalized responses. 
This turns their conversation histories into a rich repository of private information, including personal preferences, habits, and even users' secrets~\cite{iclr24contextual,fate22privacyllmsurvey,hci25selfdisclousre}. 
Like other computer systems, these applications are vulnerable to side-channel attacks, where adversaries exploit indirect information leaks from the system to steal data. 
Furthermore, the unique features of LLM-based applications, such as reasoning and memory mechanisms, provide additional attack surfaces. 
Below, we detail two prominent threats associated with LLM-powered systems: side channel attacks and information exfiltration. 

\subsection{Side Channel Attacks}

Side-channel attacks~\cite{usenix24side,lerman2011side,sidewiki} exploit indirect leakage of information through system behaviors such as timing, memory usage, and input/output patterns. 
These attacks can be especially severe in the context of LLM-based chatbots, which contain vast amounts of users' private conversation histories. 
Existing side-channel attacks against LLM-based chatbots can be categorized into three different types:

\begin{itemize}
    \item \textit{Inference Timing Attacks.}  
    Inference timing attacks target the time it takes for an LLM to generate a response. 
    To improve inference efficiency, modern LLMs are often optimized using data-dependent inference techniques, such as speculative decoding~\cite{24specinfer, nips18decoding}, where a smaller, faster ``draft'' model predicts multiple future tokens, and the larger, main model verifies them in a single step. 
    However, these optimizations introduce new vulnerabilities, as demonstrated in recent work~\cite{carlini24timing, arxiv24timing, arxiv25trading, soleiman25wiretapping}. 
    Specifically, the vulnerability arises from the number of drafted tokens that the main model accepts. 
    If many tokens are accepted, the response is faster; if most are rejected, the system slows down. 
    This acceptance rate depends on the predictability of the text. 
    Attackers can craft specific inputs to measure these timing differences and infer the predictability of a user's hidden conversation history. 
    By analyzing these server response time patterns, attackers can infer the topic or even specific characteristics of a user’s private conversations without ever seeing the actual content.
    \item \textit{Cache Timing Attacks.}
    Cache timing attacks exploit variations in how data is stored and accessed in a system's memory. In the context of LLM-based chatbots, inference services deployed on cloud resources need to handle a high volume of real-time requests while maintaining high throughput and low latency. One common optimization technique is prompt caching~\cite{mlsys23efficient}, where the attention key-value (KV) cache is reused across requests. 
    In this method, the KV cache for a prompt is stored, and if a subsequent prompt shares a matching prefix with a cached prompt, the cached KV data for the prefix can be quickly retrieved. This results in faster processing times, specifically reducing the time to generate the first response token.
    However, the use of prompt caching introduces observable variations in response times based on the private input. When a prompt matches a cached prefix, the response is faster due to the cache hit, whereas non-matching prompts result in slower response times. 
    By analyzing these timing differences, the attacker can learn the prefixes of other users' private inputs, potentially allowing them to identify or reconstruct the victim's entire prompt with high confidence~\cite{zheng2024inputsnatch, gu2025auditing, wu2025know, song2024early}.
    \item \textit{Keylogging Attacks.}
    Remote keylogging attacks focus on capturing the keystrokes entered by users during their interactions with chatbots. 
    Unlike traditional keyloggers that require local access to the user's device, recent attacks~\cite{weiss24your} show it is possible to conduct this remotely by analyzing the timing and length of network packets exchanged between the user and the chatbot. 
    By observing patterns in the size and timing of these encrypted packets, attackers can infer the length of the tokens being transmitted. 
    Leveraging the predictable structure of language, these patterns can be used to reconstruct a user's input without any access to their device.
\end{itemize}

\mypara{Privacy Implications} 
Side-channel attacks exploit indirect signals such as inference latency, cache behavior, and packet length to infer sensitive attributes, without requiring the adversary to compromise the chatbot or the user directly. 
This makes them especially dangerous, as private information can be extracted passively, often without the awareness of either party.

\subsection{Information Exfiltration}

Information exfiltration refers to the unauthorized transfer of sensitive data from one context to another.
In LLM-based (agent) applications, this occurs when attackers steal private information either through unintended leakage by the model itself or by maliciously manipulating the system to reveal user data, opening a new and highly exploitable attack surface~\cite{zhang2025llm}. 
We categorize existing approaches to information exfiltration into the following major categories.

\begin{itemize}
    \item \textit{Unintended Disclosure.} 
    LLMs often lack awareness of privacy norms and the contextual boundaries of information flows~\cite{nissenbaum04contextual}. 
    As a result, they may inadvertently disclose sensitive information to inappropriate recipients. 
    An LLM agent involved in a multi-round conversation may unintentionally repeat or expose previously shared user information, even when it is contextually irrelevant~\cite{iclr24secret, das2025disclosureauditsllmagents,shaoprivacylens,zhang2024privacy}.
    For instance, it is undesirable for an LLM assistant to reveal that ``John is talking to a few companies about switching jobs'' while drafting an email to John’s current manager, particularly without his consent. 
    This risk increases when LLMs are tasked with complex operations that involve integrating multiple sources of user data, such as combining financial, location, and preference information for personalized recommendations~\cite{shaoprivacylens}. 
    LLMs struggle to track which information is appropriate to share, which makes these disclosures particularly insidious.
    \item \textit{Leakage During Model Reasoning.}  
    Recent advances in reasoning techniques encourage LLMs to generate explicit ``thinking traces'' or intermediate reasoning steps before producing final answers~\cite{iclr23react,snell2024scaling}.  
    While this improves task performance, studies show that reasoning traces themselves may leak sensitive user data, either accidentally or via targeted prompt injections~\cite{arixv25leakprivreason}.  
    For instance, a model assisting with medical scheduling could inadvertently include a patient’s health condition in its hidden reasoning, which may later surface in outputs.  
    This creates a difficult trade-off: increasing computational effort can make an agent's final answer more cautious, but it also encourages more verbose reasoning, thereby enlarging the attack surface.
    \item \textit{Memory Leakage.}  
    To improve personalization, many commercial LLM-powered chatbots, such as ChatGPT~\cite{gpt4} and Gemini~\cite{arxiv23gemini}, have introduced long-term memory features that persist user information across sessions.  
    These memories may include personal details such as location, occupation, or user preferences, stored explicitly in textual form to improve future responses.  
    While convenient, such memory is highly sensitive and attractive to adversaries.  
    Recent studies demonstrate that attackers can exfiltrate this data via carefully designed prompt injection attacks~\cite{arxiv24exfiltration,patlan2025context}.  
    For example, malicious content embedded in a piece of code or blog post could instruct LLM to reveal stored user memories, potentially encoding them into hidden channels (e.g., URLs or snippets of code) or misleading agents to perform actions like visiting websites that acquire user information, transferring the data to a remote adversary.
    \item \textit{Insecure Tool Usage.}
    Tools refer to functions that LLM-based agents use to interact with external data or perform actions that modify the environment, such as writing files, clicking links on web pages, or generating and executing code. While the open-source community has made significant strides in developing secure Model Context Protocols (MCP) to ensure consistent and secure interactions with external data, these tools still pose substantial privacy risks. A recent study~\cite{croce2025trivial} demonstrated that MCP servers could be exploited as trojans to compromise user privacy. For example, a malicious weather MCP server, disguised as benign functionality, exploited legitimate banking tools to discover and extract user account balances. Although many vulnerabilities have been recognized~\cite{fang2025we, xing2025mcp, kumar2025mcp}, the increasing capabilities of agents with more tools at their disposal may lead to more potential privacy vulnerabilities.
    \item \textit{Compromised Execution Environment.}
    The agent's execution environment can also be manipulated to exfiltrate sensitive information. 
    For example, browser-based agents are highly susceptible to malicious prompt injections embedded in web pages~\cite{chen2025obvious, liaoeia} or triggered by pop-ups~\cite{zhang2024attacking}, leading to the leakage of private user data. 
    A recent study~\cite{kumar2025aligned} further emphasizes that agents performing GUI-based tasks are particularly vulnerable, due to the misalignment between LLM behavior in conversational settings and their behavior in agent-based, browser-use contexts.
    \item \textit{Leakge via Share Link.} 
    Users may share their conversations with a chatbot (\eg ChatGPT) through a built-in ``Share'' button on commercial chatbot platforms, allowing only those with the link to access the conversation. 
    However, it has been shown~\cite{sharegpt} that these links can be discovered through search engines (\eg Google Search). 
    This can unintentionally expose users' conversation histories to the public. 
    Furthermore, deleting a conversation from your ChatGPT history does not remove the public share link or prevent it from appearing in search engine results. 
    Recognizing the potential for privacy breaches, OpenAI has addressed this issue by providing users with the option to control whether their chats are visible in search engine results, offering more control over users' privacy.
\end{itemize}

\mypara{Privacy Implications}
Information exfiltration can occur even without the presence of an active attacker and can be unintentionally exposed through a model’s internal reasoning traces or persistent memory. 
In this way, features originally intended to improve functionality (\eg tool usage and share links), transparency (\eg reasoning), and personalization (\eg memory) become high-value attack surfaces. 
The ultimate consequence is a profound erosion of user trust, as interacting with a seemingly helpful LLM agent creates a persistent, exploitable record of their most sensitive information.

\mypara{Vulnerability Detection and Mitigation Strategies}
To counter the threat of information exfiltration, research efforts are focused on both detecting vulnerabilities and developing active defenses.
On the detection front, some works measure an LLM's capacity for privacy reasoning in ambiguous contexts to identify risks of unintended information disclosure~\cite{arxivprivreason,shao2024privacylens}. 
Another approach~\cite{ccs24airgapagent,zharmagambetov2025agentdam} examines whether agents interacting with web interfaces adhere to the principle of data minimization, introducing benchmarks to systematically evaluate their compliance.
In parallel, other efforts aim to build direct defenses against malicious attacks. 
This includes designing robust countermeasures for prompt injection attacks, which are a primary vector for information exfiltration~\cite{debenedetti2025defeating,wallace2024instruction}.
Despite these advancements, a comprehensive mitigation strategy capable of defending against the full spectrum of emerging exfiltration threats remains an open challenge~\cite{he2025comprehensive, zhang2025llm}.

\section{Privacy Risks from Malicious Use of LLMs}
\label{sec:use}

The increasingly impressive capabilities of LLMs have demonstrated remarkable potential across diverse fields, such as software engineering~\cite{arxiv24sweagent}, human behavior simulation~\cite{uist23agent}, and even assisting scientific discovery~\cite{arxiv23researchagent}. 
However, this progress presents a dual-use dilemma, as the very capabilities driving these innovations can also be misused for malicious purposes~\cite{llmscam23,llmhacker24,arxiv24autoattacker,arxiv24llminstrument,carlini2025llms}.
Specifically, LLMs amplify the risk of privacy violations in two ways:
\begin{itemize}
    \item \textit{Scaling Sophisticated Attacks.} 
    LLMs can automate and execute privacy attacks that were previously prohibitive due to their complexity or high cost. 
    By either assisting human adversaries or operating independently, they can enable privacy breaches at an unprecedented scale.
    \item \textit{Democratizing Attack Capabilities.} 
    LLMs lower the barrier for malicious actors by making powerful attack tools accessible to people with little to no expertise. 
    This ``democratization'' allows individuals with limited knowledge to launch attacks that previously required specialized skills.
\end{itemize}

In this section, we introduce two emerging privacy risks of the malicious use of LLMs: automated profile inference and automated social engineering.

\subsection{Automated Profile Inference}

Individuals constantly generate digital footprints through their online activities, encompassing activities from social media comments and posts to shared photos and videos. 
While some of this data is inherently private (\eg browse history), a vast amount of these activities (\eg posts and comments) is publicly accessible. 
However, the public availability of this information does not eliminate privacy risks. 
An adversary can aggregate these seemingly innocuous public activities to construct a detailed personal profile, a process known as profiling~\cite{profiling,profiling_book}. 
For instance, analyzing a Reddit user's most frequented subreddits could reveal their hobbies, while geotags in posted images could disclose their travel patterns or home location. 
Profiling is widely recognized as a privacy violation by legitimate privacy frameworks like GDPR\cite{gdpr}, CCPA~\cite{ccpa}, and HIPAA~\cite{hippa}.

Profiling based on unstructured and noisy data requires significant expertise and is considered too resource-intensive for large-scale privacy breaches~\cite{16doxing}. 
The emergence of LLMs fundamentally alters this landscape. 
By leveraging their sophisticated understanding and reasoning capabilities, LLMs can automate the inference process, systematically analyzing vast digital footprints to infer sensitive attributes with minimal human intervention. This automation dramatically amplifies the threat, enabling profiling attacks at an unprecedented scale.
A growing body of work has demonstrated the feasibility of LLM-driven profiling attacks~\cite{arxiv25autoprofiler,iclr24beyond,nips24imageinfer}. 
In the following, we categorize these attacks along two primary axes: (i) the data modality they target and (ii) their level of automation. 

\mypara{Profiling Across Data Modalities}
With the increasing capabilities of LLMs in understanding different data modalities, various profiling attacks have been proposed by analyzing a user's activities across multiple types of data:

\begin{itemize}
    \item \textit{Profiling from Textual Activities.} 
    Early LLM-based profiling attack~\cite{iclr24beyond} assumes that an adversary can access and scrape the public activities (\eg posts and comments) of a pseudonymous user from the Internet. 
    The adversary then instructs LLMs with prompts to infer predefined sensitive attributes (\ie eight types of PIIs), within these textual activities. 
    The results showed that powerful models like GPT-4~\cite{gpt4} can achieve performance comparable to human analysts, even when the humans have the advantage of accessing additional contextual information, which LLMs do not have.
    \item \textit{Profiling from Visual Activities.} 
    With the rise of Vision-Language Models (VLMs)~\cite{gpt4v,arxiv23gemini}, research has expanded to include profiling from images and videos, which are ubiquitous on social media platforms like TikTok and Instagram. 
    Specifically, one study~\cite{nips24imageinfer} designed carefully crafted prompts using chain-of-thought reasoning~\cite{nips22chainofthought} and automated zooming techniques to direct VLMs to focus on potentially sensitive details in the photos, thus enhancing privacy-infringing inferences. 
    Another significant privacy risk arises from directly inferring a user's possible geo-location from their pictures~\cite{geolocater,arxiv25evaluategeoinfer,arxiv25vlmgeoinfer,arxiv25imageprofiler}. 
    Research has shown that VLMs can outperform even the professional human players in GeoGuessr~\cite{geoguessr}, which raises serious concerns regarding geographic privacy. 
    However, these models are not infallible; they often exhibit significant regional biases, such as a tendency to over-predict well-known landmarks or locations heavily represented in their training data~\cite{arxiv25vlmgeoinfer}. 
\end{itemize}

\mypara{Different Levels of Automation in Profiling}
The privacy risks associated with the malicious use of LLMs depend heavily on the degree of automation involved in the attack. 
A highly automated and practical attack poses a much greater real-world privacy threat, as it reduces the need for human adversaries, making it more cost-efficient and scalable. 
We categorize existing approaches into two types: semi-automated and fully automated, depending on their level of automation:

\begin{itemize}
    \item \textit{Semi-Automated Profiling.}
    The majority of current research falls into this category, where the core inference task is automated, but significant human effort is still required for data preparation and defining attack objectives. 
    These systems are powerful in controlled settings but face two major limitations in real-world scenarios:
    (i) Reliance on curated data. Many studies~\cite{nips24imageinfer,iclr24beyond,geolocater,usenix25piiextract} focus on clean, curated textual or image data that is deliberately designed to contain sensitive information, allowing LLMs and VLMs to infer personal attributes. 
    However, in real-world scenarios, user activities are typically noisy and may not be directly related to personal attributes. 
    As a result, the performance of these semi-automated methods would likely degrade significantly when faced with raw, unfiltered activities.
    (ii) Predefined attribute targets. These attacks are typically configured to search for a fixed set of sensitive attributes (\eg age, gender, location), which assumes the adversary already knows what to profile from users. 
    However, in the real world, adversaries do not always know what sensitive attributes are present in a user's activities. 
    This lack of predefined knowledge prevents the attacker from designing specific strategies to target particular attributes, further limiting the applicability of such attacks.
    \item \textit{Fully-Automated Profiling.} 
    To address the limitations of previous approaches, recent work has focused on developing end-to-end automated profiling systems. One example is AutoProfiler~\cite{arxiv25autoprofiler}, an agent-based profiling framework that automatically scrapes, collects, and analyzes potentially sensitive activities from raw, noisy user data. By coordinating with four specialized LLM agents, AutoProfiler fully automates the process of inferring sensitive attributes. This eliminates the need for background knowledge or profiling expertise, making it highly scalable and suitable for deployment on web-scale platforms.
    Despite its weaker assumptions, the results show that the inferred attributes extend beyond PIIs, uncovering significant amounts of sensitive information. 
    The move toward full automation has profound implications. 
    It means that adversaries no longer need specialized expertise or prior knowledge to launch sophisticated, large-scale profiling attacks. 
\end{itemize}

\mypara{Privacy Implications}
Automated profiling inference can result in serious privacy breaches.
One of the most well-known risks is de-anonymization~\cite{sp08netflix,sp09network}. Study~\cite{arxiv25autoprofiler} shows that some Reddit users can be de-anonymized by inferring personal attributes from their public activities and comparing these with publicly available profiles, such as LinkedIn.
The risk of de-anonymization increases when adversaries gain access to multiple profile databases or cross-reference a user's activities to construct more comprehensive profiles. 
In addition, sensitive information extracted from these online activities can also be exploited for severe cybercrimes like doxing and cyberbullying.
We refer to~\cite{16doxing,arxiv25autoprofiler} for a deeper discussion of the consequences of exposing sensitive personal data.

\mypara{Growing Threats}
Existing attacks exploit the in-context learning (ICL) capability of off-the-shelf LLMs to perform profiling tasks.
While this approach is highly efficient and accessible, its performance could be suboptimal, as these models are not specifically designed for profiling.
For example, studies show that even state-of-the-art VLMs are outperformed in geo-location identification tasks by PIGEON~\cite{cvpr24pigeon}, an image model purpose-built for geolocation.
This trend suggests that adversaries may design specialized profiling models that surpass generic LLMs, thereby enhancing attack effectiveness and posing even more severe privacy risks.

\mypara{Challenges in Evaluation}
While various methodologies have been proposed to assess the profiling abilities of LLMs~\cite{arxiv24autoattacker,arxiv25autoprofiler,arxiv25vlmgeoinfer,iclr24beyond}, there is still no widely acknowledged benchmark to comprehensively evaluate the associated privacy risks.
This issue partially stems from a fundamental ethical dilemma: creating a robust benchmark would require a large dataset of real users' activities with labeled, sensitive, ground-truth attributes.
To address this, some researchers have proposed using synthetic datasets generated by LLM agents~\cite{nips24pai,arxiv25synpii}. 
However, the behaviors and data produced by these agents may not accurately reflect the complexities of real human activity, limiting their validity and reliability~\cite{arxiv25autoprofiler}. 
In addition, evaluation becomes more complex for fully automated systems that perform open-ended inference without predefined attribute targets.
Forcing the model to choose from a candidate list simplifies evaluation but fails to measure the model's true, unconstrained inference capabilities.
Therefore, evaluating the profiling abilities of LLMs remains an open question. 
Designing effective evaluation approaches is a critical step toward understanding and mitigating these emerging privacy threats.

\subsection{Automated Social Engineering}

A social engineering attack exploits the psychological manipulation of human behavior to extract sensitive information, gain access to personal devices, share credentials, or perform other malicious activities that compromise digital security~\cite{se_news2}. 
There are different types of social engineering, such as phishing, vishing, pretexting, and baiting.
Over the past decades, social engineering attacks have resulted in numerous incidents, causing severe financial losses and privacy breaches~\cite{social_survey1,se_news1}. 
Most social engineering attacks follow four main stages~\cite{social_survey2}:
(i) \textit{Investigation.} The attacker gathers information about the target, often from public social media, job platforms, and online sources, to identify vulnerabilities.
(ii) \textit{Planning.} Based on the gathered information, the attacker develops a strategy, selecting tactics like phishing or impersonation to exploit weaknesses.
(iii) \textit{Contact.} The attacker establishes trust with the target, persuading them to take harmful actions such as clicking a malicious link or disclosing sensitive information.
(iv) \textit{Execution.} The attacker extracts sensitive data, installs malware, or otherwise compromises the target’s system.

Social engineering attacks typically required significant human effort and expertise, and their success rates were often limited by defense mechanisms and human vigilance.
For example, phishing emails could be easily recognized by telltale signs like grammatical errors or implausible scenarios~\cite{arxiv25multiturnsocial}.
However, the advent of LLMs has introduced a new dimension to social engineering threats, which we refer to as automated social engineering. 
Unlike traditional methods, LLM-driven attacks can be personalized and executed at scale. 
These models can automate and enhance all four major stages of a social engineering attack, increasing both effectiveness and efficiency, as detailed below.

\mypara{Automated Investigation}
The purpose of this phase is to gather sufficient information about a target to personalize the attack and make it more convincing~\cite{social_survey1}.
Adversaries may directly employ automated profiling strategies (as described in the previous section) to collect personal information.
In addition, they may launch proactive information-gathering attempts by manipulating LLM-based chatbots to elicit sensitive details.
In such scenarios, a chatbot convinces the user that certain personal information is required to complete a task.
Because users often perceive LLMs as helpful assistants, they may willingly provide sensitive details, believing them to be necessary~\cite{sp25chatbot,usenix25llmmental,arxiv25discloursedefense,chi25measuredisclourse}. 
Attackers can exploit this trust by embedding hidden, privacy-invasive prompts into a chatbot’s behavior~\cite{iclr24beyond}.
For example, a chatbot tasked with creating a travel itinerary might subtly request additional personal details—such as financial information or contact numbers—under the guise of improving the service.
The risk is further amplified in multi-agent LLM systems, where multiple agents collaborate by asking for complementary pieces of information and together constructing a detailed personal profile of the victim~\cite{25aidevil}.
These LLM-based information collection strategies dramatically reduce the cost and time required for reconnaissance while producing highly detailed and actionable intelligence about targets.

\mypara{LLM-Aided Planning} 
In this stage, LLMs could serve as powerful reasoning and analysis engines to help attackers design persuasive attack strategies.
Specifically, LLMs can (i) propose tailored attack vectors—such as spear-phishing campaigns or impersonation scenarios, (ii) generate dialogue templates to sustain orchestrated interactions that gradually build trust~\cite{jailbreakse}, and (iii) dynamically adapt strategies, for example by suggesting follow-up messages when a target hesitates or fails to respond.
This capability transforms attack planning from a manual, experience-driven art into an automated process. 
Sophisticated, customized attack blueprints can be generated in minutes, removing the need for an experienced human attacker.

\mypara{LLM-Enhanced Contact}
LLMs can be exploited not only to enhance interactions through existing contact channels but also to create entirely new avenues for reaching targets.

First, LLMs enhance traditional methods like phishing by generating persuasive, context-aware emails with remarkable speed. 
Studies show an LLM could draft a highly convincing spear-phishing email in just five minutes, a task that takes a human team several hours~\cite{ibm_se1,ibm_se2,arxiv24phishingllm}.
LLMs can also sustain convincing, real-time conversations that gradually build trust.
When paired with generative deepfake technologies for images, video, or audio, impersonations become nearly indistinguishable from legitimate contacts~\cite{chatse,arxiv24defendse,arxiv25multiturnsocial}.
This allows a single attacker to maintain persistent, personalized engagement across multiple platforms and scale their outreach to thousands of potential victims simultaneously.

Second, LLMs open new avenues for attack by exploiting the growing use of chatbots for emotional and psychological support~\cite{chi25llmemotional, nature25mental}. 
In this scenario, attackers deploy malicious chatbots that impersonate trusted friends or companions to establish a deep emotional connection with a victim. 
The proliferation of third-party platforms like the OpenAI GPT Store~\cite{gptstore} and FlowGPT~\cite{flowgpt} makes it easy to distribute these deceptive chatbots to a wide audience.
Once an emotional connection is established, adversaries can manipulate victims into disclosing sensitive information, transferring money under fraudulent pretenses, or even engaging in harmful behaviors~\cite{suicidegpt}.

\mypara{LLM-Aided Execution}
Once trust is established and sensitive data is obtained, LLMs can assist attackers in carrying out malicious actions. These include:
(i) leveraging stolen credentials to gain unauthorized access to systems~\cite{arxiv25forewarned},
(ii) automating financial fraud, such as wire transfer scams~\cite{se_example}, and
(iii) orchestrating follow-on attacks, including malware distribution or pivoting to additional targets within a compromised network~\cite{mit_se}.
By reducing the need for manual effort, LLMs enable end-to-end, scalable, and highly sophisticated attack pipelines.

\mypara{Privacy and Security Implications}
Automated social engineering represents a multifaceted threat to both privacy and security.
It dramatically increases the risk of large-scale data leakage and financial loss. 
Attackers can harvest sensitive personal information, financial details, and corporate credentials with unprecedented efficiency~\cite{ibm_se1}. 
The real-world consequences are staggering; in one recent incident, fraudsters used a combination of phishing and video-based deepfake impersonation to deceive an employee into authorizing a fraudulent \$25 million transfer~\cite{se_example}.
Beyond financial loss, certain strategies exploit users’ trust or emotional reliance, inflicting psychological harm that can result in profound emotional distress.
Thus, automated social engineering not only increases the efficiency of attacks but also expands the pool of potential victims, thereby amplifying the societal impact of privacy breaches.

\mypara{Growing Threats}
With the rapid development of LLMs, automated social engineering attacks may become even more sophisticated and hard to detect. 
Multi-modal LLMs, for example, can generate coordinated text, audio, and video content to produce highly immersive impersonations that are nearly indistinguishable from authentic human interactions.
Another concern lies in the emergence of autonomous agents capable of orchestrating end-to-end attack campaigns.
Such agents could handle reconnaissance, planning, multi-turn conversations, and final exploitation without any human oversight~\cite{arxiv25multiturnsocial}.
These advancements suggest that future LLM-driven social engineering would progress beyond opportunistic scams toward coordinated, persistent, and large-scale operations capable of evading even advanced detection and defense systems.

\mypara{Vulnerability Detection and Mitigation Strategies}
Several studies have examined the capabilities of LLMs in conducting social engineering and their impact on human users~\cite{arxiv25forewarned,usenix24malla,heiding2025evaluating}.
For example, a recent work~\cite{paladin} proposed embedding trigger–tag associations into vanilla LLMs through various insertion strategies.
When the model is instructed to generate phishing emails, detectable tags are inserted into the output, enabling more effective detection of LLM-generated phishing content.
However, such safety enhancements for LLMs are limited in their real-world applicability.
Adversaries can easily bypass them by locally deploying open-source and unconstrained LLMs without these safeguards.
Thus, the challenge extends beyond detecting LLM-generated social engineering content to also identifying autonomous malicious activities carried out by LLM agents.
\section{Conclusion}
\label{sec:conclusion}

The rapid development and integration of LLMs into digital infrastructure and daily life have introduced a new frontier of privacy risks.
In this paper, we systematically examined emerging threats of LLMs across three dimensions: (i) data privacy risks across various learning stages of LLMs; (ii) privacy risks in LLM-powered applications, including side channels and information exfiltration; and (iii) malicious use of LLMs, such as automated profiling and social engineering.
We then discuss the real-world privacy implications of these threats and highlight the limitations of existing mitigation strategies.
This paper helps to illuminate the privacy risks introduced by LLMs and advocates for greater social awareness of these challenges.
We also call for research efforts that broaden their focus beyond data privacy and design new defenses to address these privacy threats.

\bibliographystyle{plain}
\bibliography{related,llm,llm_agent, data_dp,data_extraction,data_mia, weapon_profiler,weapon_social, weapon_disclosure,system_chat,system_agent,contextual}




\end{document}